\documentclass[manuscript]{aastex}
\def\H2{\hbox{H$_{2}$}}
\def\deg{^{\circ}}
\def\Lsun{\hbox{\it L$_\odot$}}

\begin{document}

\title{Very High Excitation Lines of \H2\ in  \\
the Orion Molecular Cloud Outflow}


\author{T. R. Geballe\altaffilmark{1}}
\author{M. G. Burton\altaffilmark{2}}
\author{R. E. Pike\altaffilmark{3}}

\altaffiltext{1}{Gemini Observatory, 670 N. A'ohoku Place, Hilo, HI 96720}
\altaffiltext{2}{Armagh Observatory and Planetarium, College Hill, Armagh BT61 9DG, Northern Ireland, UK}
\altaffiltext{3}{Institute of Astronomy and Astrophysics, Academia Sinica, Taipei, Taiwan, R.O.C.}

\begin{abstract}

Vibration-rotation lines of \H2\ from highly excited levels approaching the dissociation limit have been detected at a number of locations in the shocked gas of the Orion Molecular Cloud (OMC-1), including in a Herbig-Haro object near the tip of one of the OMC-1 ``fingers." Population diagrams show that while the excited \H2\ is almost entirely at a kinetic temperature of $\sim$1,800~K, (typical for vibrationally shock-excited \H2), as in the previously reported case of Herbig-Haro object HH~7 up to a few percent of the \H2\ is at a kinetic temperature of $\sim$5,000~K. The location with the largest fraction of  hot \H2\ is the Herbig-Haro object, where the outflowing material is moving at a higher speed than at the other locations.  Although theoretical work is required for a better understanding of the 5,000~K \H2, (including how it cools), its existence and the apparent dependence of its abundance relative to that of the cooler component on the relative velocities of the outflow and the surrounding ambient gas appear broadly consistent with it having recently reformed. The existence of this high temperature \H2\ appears to be a common characteristic of shock-excited molecular gas. 

\end{abstract}


\section{Introduction}

\citet[][hereafter PGBC]{pik16} recently discovered lines of \H2\ in the Herbig-Haro object 7 (HH~7) bowshock in its $K$-band (2.0-2.5~$\mu$m) spectrum, with upper level energies as high as 52,000 K.  Previously no \H2\ lines with upper level energies greater than 25,000~K had ever been observed in a shocked molecular cloud \citep[e.g.,][]{bra88,oli88,ric95}. The highest energy level of \H2\ that PGBC observed in HH~7 is just above the dissociation energy of that molecule in the ground state. All of the observed levels with energies $>$40,000~K have moderate vibrational excitation ($2~\leq ~v~\leq~5$) and high ($14~\leq~J~\leq~28$) rotational excitation; thus their line emission is not due to the often observed UV fluorescence, whose signature is observed to be the opposite of this \citep[e.g.,][]{ram93,luh96}. The highest excitation lines in HH~7 are very weak, $\sim$~1,000 times fainter than the commmonly observed bright lines of \H2, and originate in a previously unobserved hot environment in the shock-excited  gas.

The \H2\ line intensities in HH~7 were fit very well by PGBC using a two-temperature model, with 98.5\%\ of the gas at $T$~=~1,800~K (a typical observed temperature for post-shock vibrationally excited molecular gas) and 1.5\%\ at $T$~=~5200~K. PGBC discussed various interpretations of the fit and tentatively concluded that the hotter component arises in \H2\ that has just re-formed on and been ejected from grains following dissociation of \H2\ by the shock. The well-defined high temperature of $\sim$5,000 K is puzzling. Current understanding of newly re-formed \H2\ suggests that initially it might not have a well-defined kinetic temperature and that in any event its kinetic temperature should not be restricted to values near 5,000~K \citep[e.g.,][]{dul86,dul93}. Indeed, the hot \H2\ ought to be cooling rapidly and so should not be associated with a single well-defined temperature.

Shock fronts in molecular clouds typically are created when outflowing gas from pre-main sequence stars encounters gas in the ambient molecular clouds out of which the stars formed and/or encounters previously ejected but slower moving gas and swept up material from the cloud.  In a pure hydrodynamic shock \H2\ is dissociated when collisions involving it occur at speeds exceeding 20--24~km~s$^{-1}$ \citep{hol77,kwa77,lon77}. In many instances the difference between the velocities of the outflow and ambient cloud far exceeds this limit; yet in many such cases (including HH~7) strong \H2\ lines with broad velocity profiles are seen \citep[e.g,][]{nad79}; thus it is apparent that much of the shocked \H2\ survives. Continuous shocks, in which the ambient gas is gradually accelerated by ions in a magnetic field, have long been the leading candidate to explain the survival of the \H2 \citep{dra80,che82,dra82}. The small percentage of highly excited \H2\ in HH~7 suggests that C-shocks are not total shields against collisional dissociation of \H2, however.

\begin{figure}[http]
\begin{center}
\includegraphics[width=0.6\textwidth]{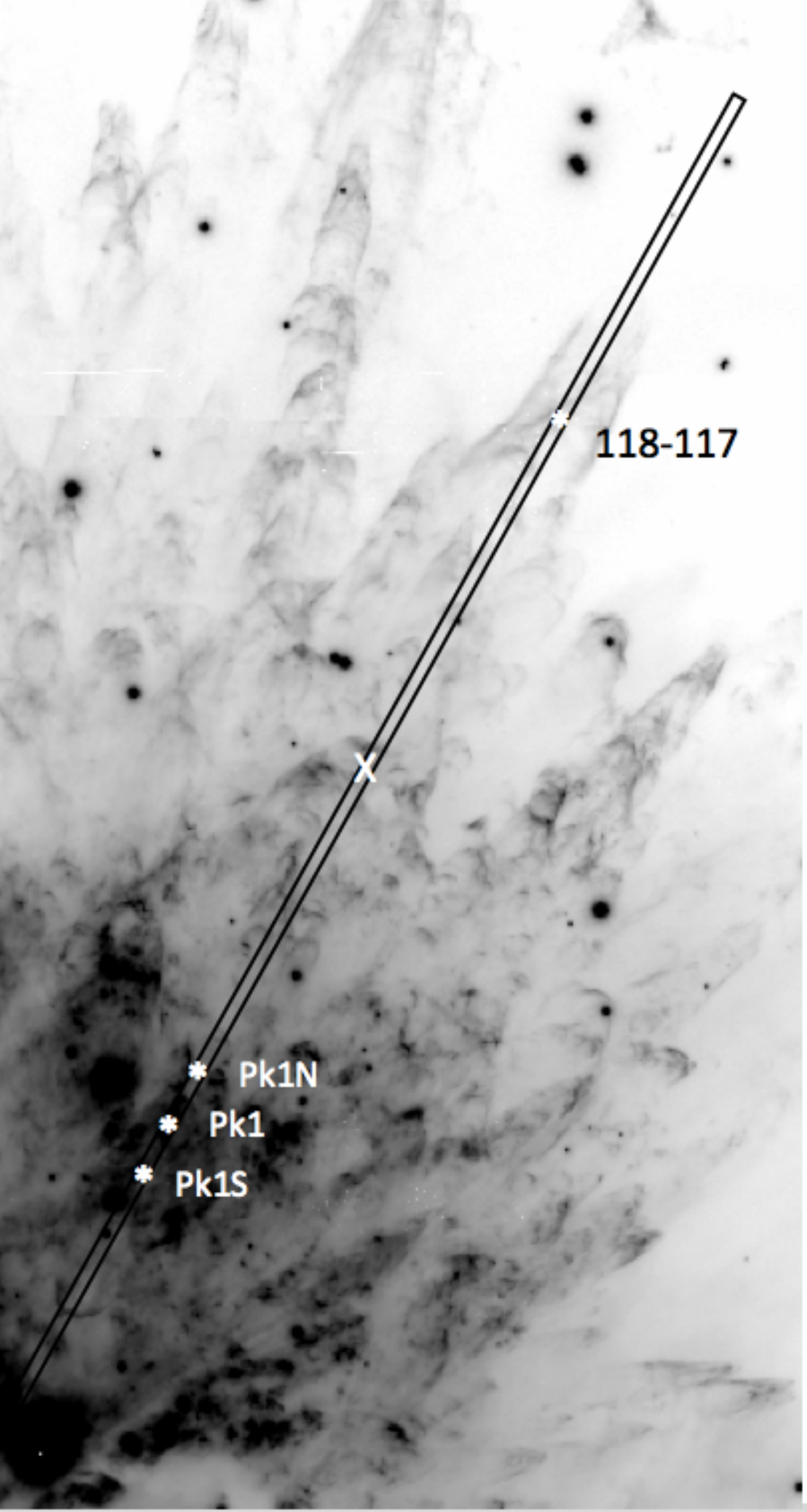}
\caption{Image of a portion of the OMC-1 outflow in the \H2\ 1--0 S(1) line \citep{bal15} with the GNIRS slit superimposed. The midpoint of the 99-arcsecond long slit, indicated by the cross, is RA = 5:35:12.7, Dec = -5:21:41.25 (2000) and the position angle of the slit is -28.72$\deg$. The centers of the four segments of the slit from which spectra have been extracted are indicated by asterisks.}
\label{slitpos}
\end{center}
\end{figure}

Whatever the explanation for the 5,000~K \H2\ in HH~7, it is important to determine if similarly hot \H2\ is common elsewhere in molecular gas that is excited by high velocity shocks. To begin to address this question, we have obtained $K$-band spectra of the brightest known region of line emission from shocked \H2, the energetic outflow in the Orion Molecular Cloud (OMC-1), in which the above mentioned dissociation Òspeed limitÓ is greatly exceeded \citep{nad79}.  Using the Frederick C. Gillett Gemini North telescope and its near-infrared spectrograph, GNIRS, with its long slit, we have simultaneously sampled a range of shocked gas environments in this cloud, from the region of brightest line intensity and broad line emission near Peak 1 \citep{bec78} through one of the many Herbig-Haro objects in OMC-1 produced by high speed ``bullets" \citep{all93}.

\begin{figure}[http]
\begin{center}
{\includegraphics[width=0.55\textwidth]{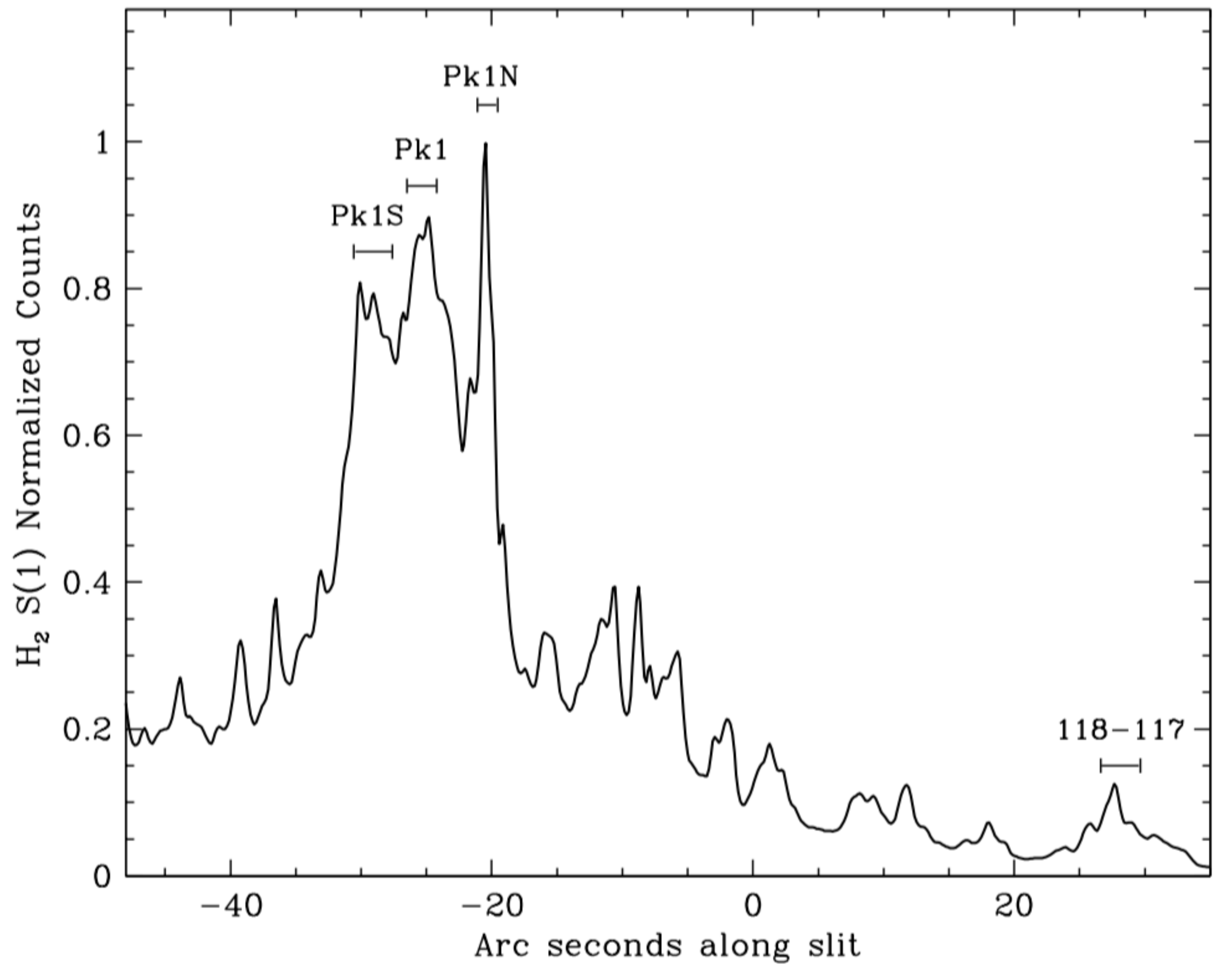}}
\caption{Intensity of \H2\ 1--0 $S$(1) line emission along the GNIRS slit shown in Fig.\ref{slitpos}. The zero offset position corresponds to  RA  =  5:35:12.7, Dec = -5:21:41.25 (2000). Positive offsets are to the north-northwest. The regions from which spectra were extracted are indicated.}
\label{slitcounts}
\end{center}
\end{figure}

\section{Observations and Data Reduction}

Long-slit spectra covering 1.9--2.5~$\mu$m were obtained at Gemini North on UT 2015 November 6 and 8 for program GN-2015B-FT-3. The 0.45~$\times$~99 arcsecond slit and the 32 lines/mm grating in GNIRS were employed, providing a resolution of 0.0018~$\mu$m, corresponding to a resolving power of $\sim$1200. This is considerably lower resolution than was used by PGBC to observe HH~7, and is insufficient to even partially resolve the \H2\ velocity profiles in OMC-1 that have been observed by others. However, it is sufficient to separate the wavelengths of some of the critical (and very weak) high excitation \H2\ lines from neighboring stronger \H2\ lines, and from recombination lines from the foreground H~II region. 

Figure~\ref{slitpos} is a portion of an image of the \H2\  2.1218-$\mu$m 1--0 S(1) line emission, kindly supplied by J. Bally, on which the location of the GNIRS slit, oriented 28.72 degrees west of north, is superimposed. Using an offset from the bright star $\theta^{1}$Ori C the approximate midpoint of the sit was positioned half way between OMC-1 Peak 1, the region of brightest \H2\ line emission, and one of the brighter OMC-1 Òbullets,Ó approximately 55 arcseconds to the north-northwest, referred to as M42 HH120--114 by \citet{ted99} and Finger 6 by \citet{bal17}. Observations were obtained in the standard ABBA mode with the B position 90 arc-seconds to the west, where \H2\ line emission from shocked gas is much weaker. The total on-source plus off-source exposure time was 48 minutes on each night. 

Data reduction followed standard procedures, utlilizing flat-fielding, ratioing by the spectrum of the telluric standard star HIP 22189 for removal of telluric absorption lines and flux calibration, and observing the spectrum of an argon arc lamp for wavelength calibration. All of these measurements were obtained near-simultaneously with the spectra of OMC-1. Figure~\ref{slitcounts} shows the strength of the 1--0 $S$(1) line along the slit. After examination of this plot, four regions of bright \H2\ line emission, shown in  Fig.~\ref{slitpos} and described in Table~\ref{regions}, were selected for detailed study. The three brightest of these are associated with Peak 1; the fourth lies in HH120--114 / Finger 6, approximately 10 arcseconds south-southwest of the tip of the shock as vewed in the  Fe~II 1.64-$\mu$m image of \citet{bal15}. Hereafter we refer to this location in the nomenclature of \citet{doi02} as 118-117. For each of these regions the reduced spectra from the two nights, which were consistent with one another to within the noise, were aligned in wavelength and averaged to produce final spectra. The wavelength calibration is accurate to 0.0003~$\mu$m.

\begin{table}[http]
\label{regions}
\caption{Locations of Extracted Spectra}
\begin{center}
\scriptsize
\begin{tabular}{|c|c|c|c|}
\hline
Name & Center RA \& Dec & Area & $A$(2.2$\mu$m) \\
  & (2000) & arcsec$^{2}$ & mags \\
 \hline
Pk 1 S &  5:35:13.63~~-5:21:06.7 & 0.45~$\times$~ 3.00  & 1.14 \\
Pk 1 & 5:35:13.52~~-5:22:03.5 & 0.45~$\times$~ 2.25 & 1.10 \\
Pk 1 N & 5:35:13.35~~-5:21:51.1 & 0.45~$\times$~ 1.50 & 1.39  \\
118--117 & 5:35:11.80~~-5:21:16.7 &  0.45~$\times$~ 3.30 & 0.22 \\
\hline
\end{tabular}
\end{center}
\end{table}

\begin{figure}[http]
\begin{center}
{\includegraphics[width=1.06\textwidth]{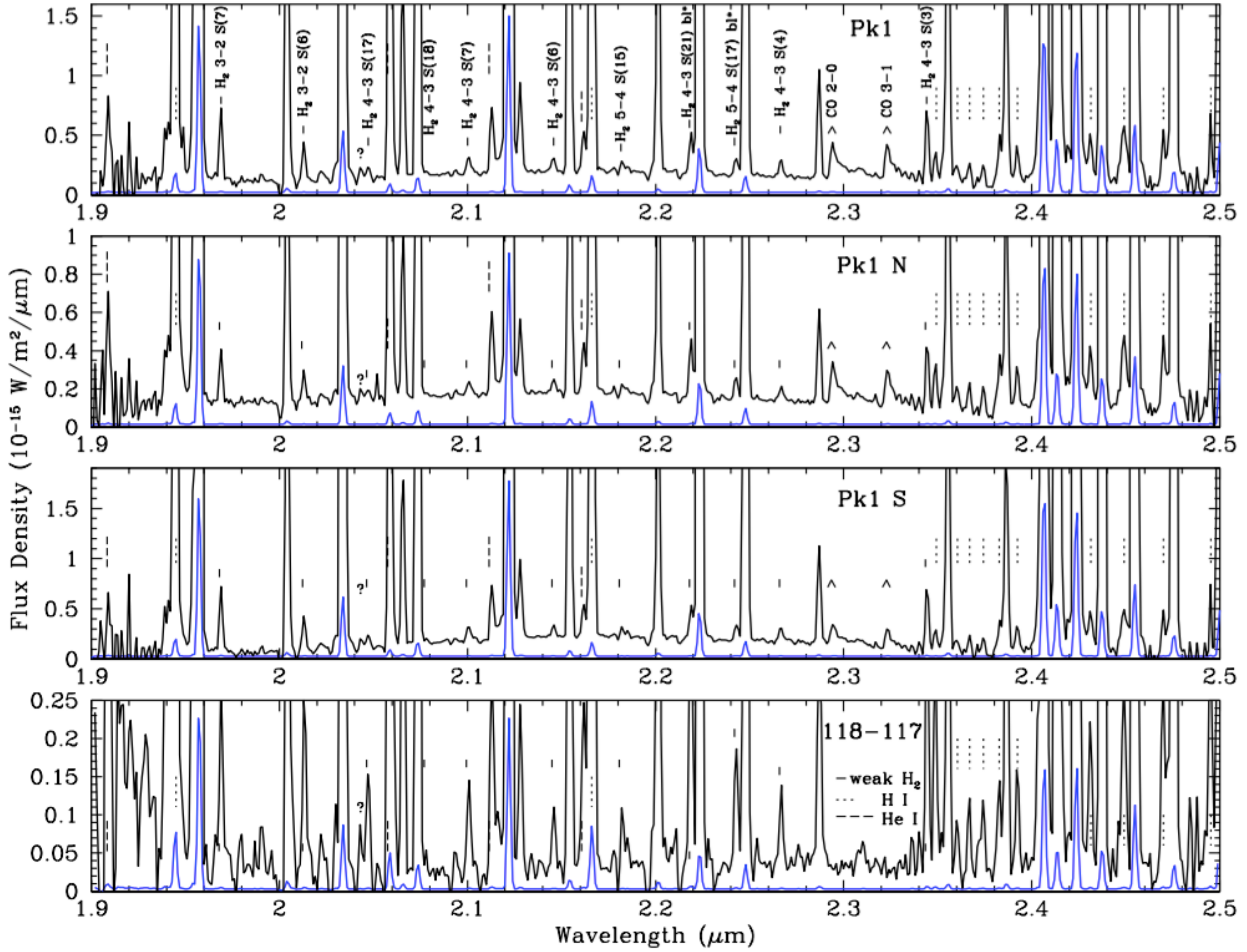}}
\caption{$K$-band continuum-subtracted spectra of four regions in OMC-1. \H2\ 1--0 2.1218~$\mu$m $S$(1) line peaks are full scale in blue spectra. Spectra in black are magnified 100 times; the flux density scale applies to them. Strong \H2\ lines are unlabeled; weak \H2\ lines are indicated by tick marks and labeled at top.  Lines of other species are indicated according to the legend at bottom. Blends of weak \H2\ lines with [Fe III] lines are indicated by asterisks.}
\label{allspec}
\end{center}
\end{figure}

\begin{table}[http]
\caption{Detected Lines}
\label{lines}
\begin{center}
\tiny
\begin{tabular}{|c|c|c|c|c|c|c|c|c|}
\hline
Rest $\lambda$ & Species & Transition & Upper Level & Pk1 Flux & Pk1N Flux & Pk1S Flux & 118--117 Flux & Notes \\
vac. $\mu$m &  & & Energy (K) & 10$^{-18}$W m$^{-2}$ & 10$^{-18}$W m$^{-2}$ & 10$^{-18}$W m$^{-2}$ & 10$^{-18}$W m$^{-2}$ &  \\
\hline
1.9095 & He I & 4-3 & & yes & yes & yes & yes & yes = detected \\
1.9447 & H I & 8--4 & & yes & yes & yes & yes & \\ 
1.9576 & \H2\ & 1-0 $S$(3) & 8,365 & 301 & 190 & 344 & 51.0 &  \\ 
1.9692 & \H2\ & 3-2 $S$(7) & 23,070 & 1.10 & 0.54 & 1.16 & 0.42 & \\
2.0041 & \H2\ & 2-1 $S$(4) & 14,764 & 7.5 & 3.66 & 7.38 & 2.3 &  \\
2.0130 & \H2\ & 3-2 $S$(6) & 21,912& 0.73 & 0.35 & 0.74& 0.53 & \\
2.0334 & \H2\ & 1-0 $S$(2) & 7,585 & 110 & 63.3 & 126 & 17.5 & \\
2.0430 & uid & & & 0.14 & 0.08 & 0.17 & 0.04 & \\
2.0475 & \H2\ & 4-3 $S$(17) & 42,022 & 0.15 & 0.08 & 0.23 & 0.32 & \\
2.0587 & He I & $2^{1}$P - $2^{1}$S & & yes & yes & yes & yes & \\
2.0656 & \H2\ & 3-2 $S$(5) & 20,857 & 3.60 & 1.83 & 3.62 & 1.40 & \\
2.0735 & \H2\ & 2-1 $S$(3) & 13,890 & 27.6 & 15.5 & 29.6 & 6.90 & \\
2.0781 & \H2\ & 4-3 $S$(18) & 43,614 & 0.06 & 0.04 & 0.07 & 0.04& \\ 
2.1004 & \H2\ & 4-3 $S$(7) & 27,202 & 0.40 & 0.22 & 0.43 & 0.27 & \\
2.1126 & He I & 4-3 & & yes & yes & yes & yes & \\
2.1218 & \H2\ & 1-0 $S$(1) & 6,952 & 315 & 188 & 374 & 46.2 & \\
2.1281 & \H2\ & 3-2 $S$(4) & 19,912 & 1.76 & 0.90 & 1.75 & 0.51& \\
2.1460 & \H2\ & 4-3 $S$(6) & 26,615 & 0.27 & 0.17 & 0.25 & 0.23 & \\ 
2.1542 & \H2\ & 2-1 $S$(2) & 13,151 & 12.1 & 6.21 & 11.7 & 2.75 & \\
2.1615 & He I & 7-4 & & yes & yes & yes & yes & \\
2.1661 & H I & 7-4 & & yes & yes & yes & yes & \\ 
2.1818 & \H2\ & 5-4 $S$(15) & 42,379 & 0.16 & 0.08 & 0.17 & 0.15 & \\
2.2010 & \H2\ & 4-3 $S$(5) & 25,624 & 1.51 & 0.83 & 1.60 & 0.52 & blend; see PGBC \\
2.2014 & \H2\ & 3-2 $S$(3) & 19,087 & 4.54 & 2.51 & 4.82 & 1.56 & blend; see PGBC \\
2.2184 & [Fe III] & & & 0.33 & 0.28 & 0.33 &0.41 & blend; see text \\
2.2196 & \H2\ & 4-3 $S$(21) & 48,345 & 0.33 & 0.28 & 0.33 & 0.41 & blend; see text \\
2.2235 & \H2\ & 1-0 $S$(0) & 6,472 & 80.6 & 49.0 & 96.7 & 10.6 & \\
2.2427 & [Fe III] & & & 0.11 & 0.09 & 0.11 & 0.14 & blend; see text \\
2.2433 & \H2\ & 5-4 $S$(17) & 45,275 & 0.23 & 0.15 & 0.28 & 0.10 & blend; see text  \\
2.2477 & \H2\ & 2-1 $S$(1) & 12,551 & 29.9 & 17.8 & 32.9 & 6.83 & \\
2.2666 & \H2\ & 4-3 $S$(4) & 24,734 & 0.27 & 0.19 & 0.38 & 0.24 & \\
2.2870 & \H2\ & 3-2 $S$(2) & 18,387 & 1.88 & 1.01 & 2.03 & 0.70 & \\
2.2935 & CO & 2-0 bh & & yes & yes & yes & no & \\
2.3227 & CO & 3-1 bh  & & yes & yes & yes & no & \\
2.3445 & \H2\ & 4-3 $S$(3) & 23,955 & 1.17 & 0.83 & 1.39 & 0.80 & \\
2.3488 & H I & 29-5 & & yes & yes & yes & yes & \\
2.3555 & \H2\ & 2-1 $S$(0) & 12,095 & 7.1 & 4.8 & 7.45 & 1.50 & \\
2.3604 & H I & 27-5 & & yes & yes & yes & yes & \\
2.3669 & H I & 26-5 & & yes & yes & yes & yes & \\
2.3744 & H I & 25-5 & & yes & yes & yes & yes & \\
2.3828 & H I & 24-5 & & yes & yes & yes & yes & \\
2.3863 & \H2\ & 3-2 $S$(1) & 17,819 & 4.77 & 2.91 & 4.66 & 1.48 & \\
2.3922 & H I & 23-5 & & yes & yes & yes & yes & \\
2.4066 & \H2\ & 1-0 $Q$(1) & 6,149 & 308 & 193  & 375 & 35.9 & \\
2.4133 & \H2\ & 1-0 $Q$(2) & 6,472 & 108 & 64.5 & 125 & 12.1 & \\
2.4237 & \H2\ & 1-0 $Q$(3) & 6,952 & 274 & 173 & 328 & 33.8 & \\
2.4311 & H I & 20-5 & & yes & yes & yes & yes & \\
2.4375 & \H2\ & 1-0 $Q$(4) & 7,585 & 90.0 & 55.6 & 107 & 12.1 & \\
2.4487 & H I & 19-5 & & yes & yes & yes & yes & \\
2.4547 & \H2\ & 1-0 $Q$(5) & 8,365 & 128 & 76.5 & 158 & 23.1 & \\
2.4669 & H I & 18-5 & & yes & yes & yes & yes & \\
2.4755 & \H2\ & 1-0 $Q$(6) & 9,286 & 46.3 & 26.7 & 49.1 & 7.2 & \\
\hline
\end{tabular}
\end{center}
\end{table}

\section{Results}

\subsection{Detected Lines}

The final reduced spectra of the four regions are shown in Fig.~\ref{allspec}. Note that the the solid angles that they cover are not identical; they vary by over a factor of two (see Table~\ref{regions}). In each spectrum the continuum, which was determined by fitting a spline through wavelengths devoid of line emission, is subtracted. The strongest \H2\ lines are much weaker at 118--117 (bottom panel of Fig.~\ref{allspec}) than at the other positions; hence the spectrum as displayed there appears noisier. 

A list of detected lines, including fluxes measured for the lines of \H2\ (by numerical integration) is given in Table~\ref{lines}. More than half of the detected lines are from \H2. Recombination lines of H~I (Brackett~$\gamma$ and high $n$ lines from the Pfund series) are also present, as are a few lines of He~I. We also tentatively identified two forbidden lines of Fe~III, present in spectra by \citet{you16} of other HH objects in OMC-1, and which are blended with weak high excitation \H2\ lines. The latter three sets of lines presumably arise in the Orion Nebula H~II region and are only partially removed by subtraction of the line emission at the nearby ÒskyÓ position, which also lies within the nebula. In addition, near 2.3~$\mu$m CO 2--0 and 3--1 band head emission is observable at the three locations near OMC-1 Peak 1. This is likely to be scattered emission that originated in the hot and dense disk surrounding the Becklin-Neugebauer object \citep{sco79} and possibly in disks of other other embedded young stellar objects within OMC-1.

The relative intensities of the \H2\ lines in the three spectra near Peak 1 are nearly identical. The relative intensities in spectrum of 118--117 are strikingly different in two ways. First and most notably, the weak \H2\ lines that can only be seen in the highly magnified spectra, are several times stronger relative to the strong lines at 118-117 than at the Peak 1 positions. Their fluxes are in fact comparable to their fluxes in the Peak 1 spectra, unlike those of the strong lines. The weak lines all arise from highly excited energy levels, as can be seen in Table~\ref{lines}. This hints at the existence of a high temperature component at 118--117 and/or a larger fraction of high temperature \H2\ at 118--117 than at the other three locations.  Second, at 118--117 the bright 1--0 $Q$-branch lines at 2.40--2.48~$\mu$m are much weaker relative to the bright 1--0 $S$(1-3) lines in the short wavelength portion of the spectrum.  This implies significantly lower extinction to the shocked \H2\ in 118--117 than at Peak 1.

\begin{figure}[http]
\begin{center}
\includegraphics[angle=-90,origin=c,width=.44\textwidth]{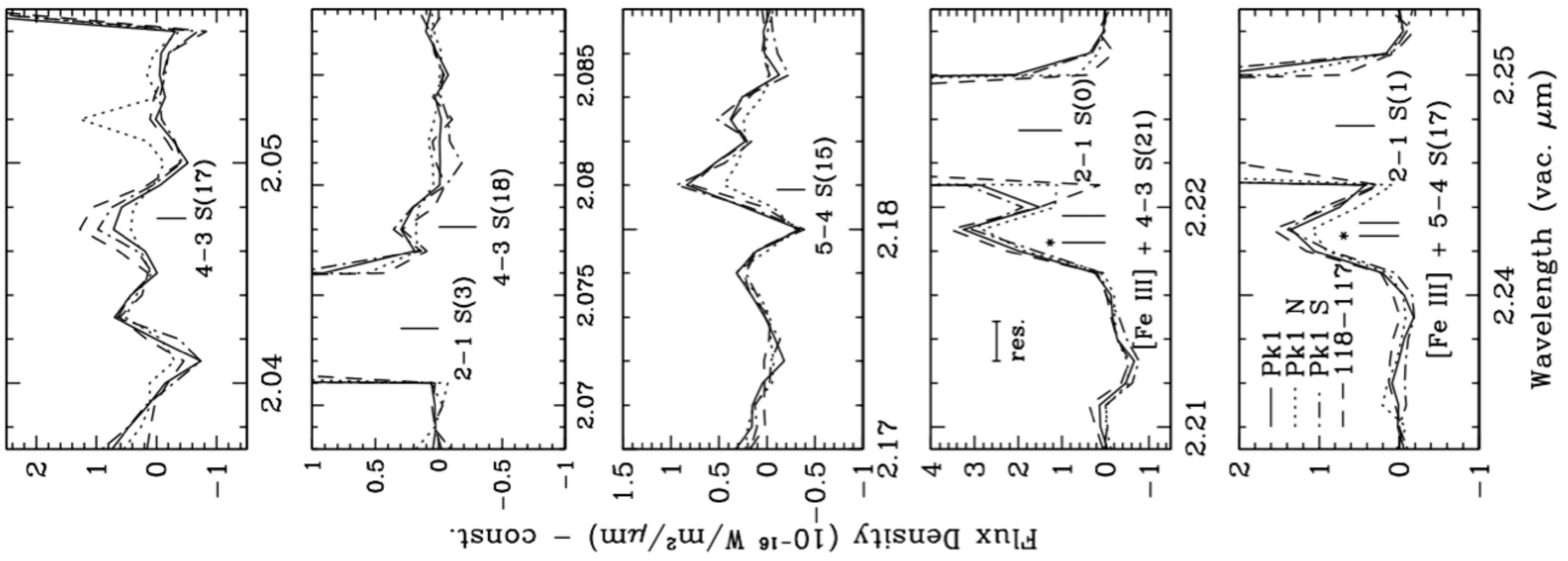}
\caption{Spectra near the wavelengths of five weak, high excitation \H2\ lines, shown for all four regions. Rest wavelengths of these lines and of nearby stronger \H2\ lines are indicated by vertical lines. Forbidden lines of Fe III are indicated by vertical lines with asterisks.}
\label{highexc}
\end{center}
\end{figure}

\subsection{The Highest Excitation Lines}

As mentioned earlier, due both to the lower spectral resolution and to the presence of nearby line emission from other species, many of the \H2\ lines from the highest energy levels that were seen by PGBC toward HH~7 were not detected at the four positions in OMC-1.  Spectra in the vicinity of the five lines with upper energy levels exceeding 40,000~K that were detected are shown in Fig.~\ref{highexc} for all four regions. Three of the lines are well resolved from neighboring lines, but the  4--3 $S$(21) and 5--4 $S$(17) lines in the lower two panels are blended with forbidden Fe III lines; the latter line badly so. We estimate the fluxes in these \H2\ lines as follows.  The Fe~III 2.2184-$\mu$m and \H2\  2.2196-$\mu$m lines are shifted in wavelength from the center of the 2.219~$\mu$m emission feature by the same amount and thus probably contribute approximately equally to the feature. We thus have set the fluxes in the 4--3 $S$(21) line to half of the measured fluxes of the 2.219-$\mu$m feature. The spectra from \citet{you16} indicate that the 2.2427-$\mu$m line of Fe~III is about one-third the strength of the 2.2184-$\mu$m line of Fe III. We subtract these values from the 2.243-$\mu$m feature to derive the strengths of \H2\ 5--4 $S$(17) line. The fluxes of the two \H2\ lines, which are listed in Table~\ref{lines}, have higher uncertainties than the other weak lines. We have allowed for this by tripling their uncertainties. However, their effects on the model fits to the \H2\ level populations are minor.  

\subsection{Fluorescent Contribution}

\citet{luh94} detected faint highly extended emission in the Orion Molecular Cloud from the 1--0 $S$(1) and 6--4 $Q$(1) lines of H$_{2}$, which they identify as  ultraviolet induced fluorescence. This emission is superimposed on the emission from the collisionally shocked gas. The extent of the fluorescent line emission (at least 2 square degrees) is considerably greater than the of the line emission from shocked gas ($\sim$10 square arc-minutes). \citet{luh94} estimate a fluorescent luminosity of 34 ~\Lsun\ in the S(1) line alone. Averaged over the emission region the surface brightness of this fluorescent line is 150 times fainter than that of 118--117, the faintest region of shocked H$_{2}$ that we have examined. The small scale intensity distribution of the fluorescent emission is unknown, but the distribution is likely to be fairly smooth, unlike that of the line emission from the shocked gas. Thus subtraction of the fluorescent signal in the ``sky" spectrum, obtained only 1\farcm5 distant is probably nearly complete. Moreover, the high excitation lines that most tightly constrain our analysis are emitted from rotation levels of 17 or higher. Radiative excitation of cold or even warm H$_{2}$ is unlikely to populate those levels. For all of these reasons we are confident that residual fluorescence makes a negligible contribution to the spectra of H$_{2}$ reported here and thus can be safely ignored in our analysis.

\begin{figure}[http]
\begin{center}
{\includegraphics[width=1\textwidth]{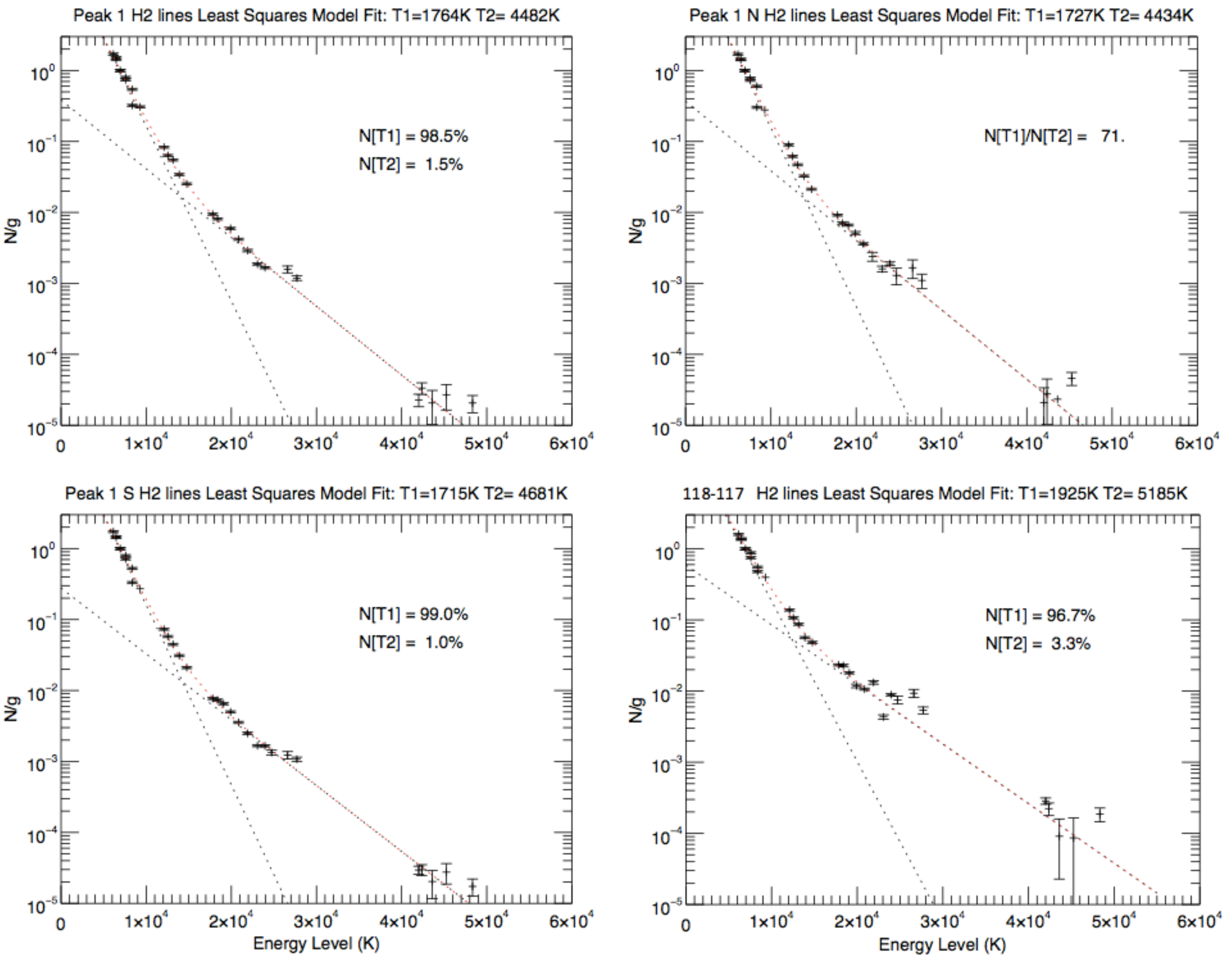}}
\caption{Level column densities, divided by their degeneracies plotted as a function of level energy for the shocked H$_2$ line emission. The y-axes are normalized to unity for the $(v,J) = (1,3)$ level (corresponding to the upper level of the 1--0 $S$(1) line). Error bars correspond to 1$\sigma$ values of 0.03 for the intensities listed in Table~\ref{lines}, for all but two weak lines for which they are 0.09 (see text). The curved dashed lines, the sums of the two straight dashed lines, are the best two-temperature LTE fits, for an ortho-to-para ratio of 3.  Note the significant difference in the temperature of the hotter component at 118--117 compared to its values in the Peak 1 region, and the much larger percentage of hot gas at 118--117.}
\label{h2models}
\end{center}
\end{figure}

\section{Analysis of the \H2\ Spectrum and Results}

Our method of analysis was the same as described in detail by PGBC. Briefly, after correcting the line fluxes for extinction (values given in Table~\ref{regions}), which was determined from the intensity ratios of the 1--0 $S$(1) and $Q$(3) lines, \H2\ level column densities were calculated using decay rates from \citet{wol98} and line frequencies from \citet{tur77}. The resultant population diagrams are shown in Fig.~\ref{h2models}. 

As in the case of HH~7, each of the four population diagrams is clearly indicative of gas at two distinct temperatures. The diagrams also demonstrate via the presence of significant populations in high rotational states that the higher temperature \H2\ is not radiatively (UV) excited. We therefore modelled the population diagrams for two-temperature Boltzmann distributions as described in PGBC, again using the Levenberg--Marquard minimum-$\chi^{2}$ fitting routine and setting the maximum signal-to-noise ratio at 30 to avoid being constrained by the brightest lines, for which the uncertainties in flux are limited by systematic errors rather than random noise. The best fitting models are plotted in Fig.~\ref{h2models}, along with the straight lines corresponding to the individual temperature components of which the model curves consist. 

As was found for HH~7, at each location we examined in the OMC-1 shocked gas, a hot component of vibrationally excited \H2\ is present at the few percent level and at a temperature similar to that of HH~7.  At the Peak 1 locations the hot components have similar temperatures of 4,500~K to within the uncertainties (several hundred Kelvins), whereas the temperature of the hot component at 118--117 is $\sim$700~K higher and is essentially identical to the value found in HH~7. The temperature of the cooler bulk component at 118--117 is also higher than elsewhere in OMC-1, by roughly 200~K, and may be marginally higher than HH~7.  Most interestingly, 118--117 also is significantly different from the Peak 1 positions and from HH~7 in that its percentage of hot \H2\ is 2--3 times greater than at Peak 1 and is twice that of HH~7. Although the lines from the most highly excited levels (upper level energies in the range 40,000--50,000~K), whose strengths are most uncertain, are in part responsible for this conclusion, examination of  the population diagrams (Fig.~\ref{h2models}) shows that even in the range 20,000--30,000K the difference between 118--117 and the Peak 1 locations is obvious.

\citet{goi15} have performed a similar anaysis of far-infrared (pure rotational) line emission from CO, H$_{2}$O and OH in the core of the Orion Molecular Cloud including the region near Peak 1. They too find multiple temperature components. The far-infrared line emission is dominated by molecular gas at much lower temperatures than that responsible for the H$_{2}$ vibration-rotation band line emission. However, at Peak 1 weak CO line emission is present  from levels as high as $J$=48, 6,458~K above the ground state, comparable to the upper level energies of the strong 1--0 $S$(0) and $S$(1) lines of \H2. Their analysis indicates that this emission arises predominantly from gas at $\sim$2500~K. That gas likely also produces the bulk of the H$_{2}$ vibration-rotation band emission.

\section{Discussion}

The results of this paper, combined with the previous result of PGBC, strongly indicate that the presence of a small amount of $\sim$5,000~K \H2\ is a common feature of molecular gas that is shocked by high velocity outflows and for which the temperature of most of the vibrationally excited \H2\ does not exceed the typically observed value of $\sim$2,000~K. Despite the large scatter of the fluxes of the weak, high excitation lines in the population diagrams, it appears that the temperature difference between the high temperature \H2 (5,000--5,200~K) in HH~7 (PGBC) and 118-117 and the high temperature \H2 (4,400--4,700~K) at the Peak~1 locations is real.

PGBC tentatively concluded that the \H2\ at $\sim$5,000~K has recently re-formed on grains after some of the \H2\, either in the outflow or in the ambient gas, has been collisionally dissociated by the shock. However, as they discussed (see PGBC for details and references), how and why such gas should have a well-defined temperature is unclear. Some theoretical work on the energy level distribution of reformed \H2\ that is just ejected from the dust grains on which it formed has indicated that is level distribution is unlikely to be thermal \citep{dul86,dul93}.  More recent laboratory experiments, summarized in \citet{wil07}, found that the internal energy of the \H2\ after ejection from graphitic surfaces is less than 40\%\ of the \H2\ binding energy of 4.5~eV. As the observed temperature of $\sim$5,000~K, for the hot \H2\ corresponds to 0.5~eV, the laboratory and astronomical results are consistent. The \H2\ level populations in the laboratory experiment were not measured, however. Comparison of future such measurements with the astronomical data would be of great interest.

Even if the initial distribution of the reformed \H2\ is thermal and at a temperature well above that of its surroundings, one might naively expect it to cool rapidly radiatively and/or collisionally and not to be characterized by a single well-defined temperature, but rather a range of temperatures. Modeling the time dependence of the energy level distribution starting from a variety of initial conditions and in a variety of environments could be illuminating, as would much more accurate astronomical measurements of the relative fluxes of some of the most highly excited lines.

\citet{ted99} derived a shock speed of 120 km~s$^{-1}$ near 118--117 by combining velocity profiles with proper motion measurements. Proper motion measuremements near Peak 1 are unavailable; however, examination of Fig.~5 of \citet{bal15} suggests that motions are relatively low there.  The velocity profiles at Peak 1 \citep{nad79,nad82}, may then be indicative of actual wind speeds there. If so, they show that only a small fraction of the gas is moving at speeds approaching 120 km~s$^{-1}$.  Because higher velocity outflows impinging on ambient cloud material are more likely to lead to increased dissociation of \H2, qualitatively one would expect a higher percentage of the \H2\ to be dissociated at 118--117 than at Peak~1 and likewise a higher percentage of re-formed \H2\ there, as is apparently observed. However, it is unclear why the temperature of the reformed \H2 would be higher at those locations, as is apparently the case.

Finally, it would be of interest to observe the spectrum of \H2\ at other locations in the OMC-1 outflow where speeds are considerably higher to determine if the fraction of high temperature \H2\ is even greater than toward 118--117.   Likewise, it would be  of interest to observe the spectrum in lower velocity molecular shocks including in regions in which J-shocks, rather than C-shocks are thought to be the dominant interaction between outflow and cloud.

\begin{acknowledgements}

The data presented here were obtained at the Gemini Observatory, which is operated by the Association of Universities for Research in Astronomy, Inc., under a cooperative agreement with the NSF on behalf of the Gemini partnership: the National Science Foundation (United States), the National Research Council (Canada), CONICYT (Chile), Ministerio de Ciencia, Tecnolog'a e Innovacion Productiva (Argentina), and Ministerio da Ciencia, Tecnologia e Inovacao (Brazil). We are grateful to the referee, J. Bally, for a number of helpful comments. 

\end{acknowledgements}

\end{document}